\documentclass[%
 reprint,
 superscriptaddress,
preprintnumbers,
nofootinbib,
 amsmath,amssymb,
 aps,
prl,
]{revtex4-1}

\usepackage{amsmath}
\usepackage{amssymb}
\usepackage{amsthm}
\usepackage{MnSymbol}
\usepackage{nicefrac}
\usepackage{amsfonts}
\usepackage{dsfont}

\usepackage[markup=nocolor]{changes}
\usepackage{natbib}

\usepackage{graphicx}

\usepackage[normalem]{ulem}

\usepackage{dcolumn}
\usepackage{bm}

\usepackage{color}
\usepackage[usenames,dvipsnames,svgnames,table]{xcolor}
\usepackage{slashed}
\usepackage{empheq}
\newcommand{\bea}{\begin{eqnarray}}
\newcommand{\eea}{\end{eqnarray}}
\newcommand{\be}{\begin{equation}}
\newcommand{\ee}{\end{equation}}
\usepackage[normalem]{ulem}

\definecolor{mygreen}{rgb}{0, 0.6, 0}

\begin{document}

\title{Mass difference for charged quarks from asymptotically safe quantum gravity}
 
 \author{Astrid Eichhorn}
   \email{a.eichhorn@thphys.uni-heidelberg.de}
\affiliation{Institut f\"ur Theoretische
  Physik, Universit\"at Heidelberg, Philosophenweg 16, 69120
  Heidelberg, Germany}
\author{Aaron Held}
\email{a.held@thphys.uni-heidelberg.de}
\affiliation{Institut f\"ur Theoretische
  Physik, Universit\"at Heidelberg, Philosophenweg 16, 69120
  Heidelberg, Germany}

\begin{abstract}
We propose a scenario to retrodict the top and bottom mass and the Abelian gauge coupling from first principles in a microscopic model including quantum gravity. In our approximation, antiscreening quantum-gravity fluctuations induce an asymptotically safe fixed point for the Abelian hypercharge leading to a uniquely fixed infrared value that is observationally viable for a particular choice of microscopic gravitational parameters. The unequal quantum numbers of the top and bottom quark lead to different fixed-point values for the top and bottom Yukawa under the impact of gauge and gravity fluctuations. This results in a dynamically generated mass difference between the two quarks. To work quantitatively, the preferred ratio of electric charges of bottom and top in our approximation lies in close vicinity to the Standard-Model value of $Q_b/Q_t =-1/2$.
\end{abstract}

\pacs{Valid PACS appear here}

\maketitle
\begin{figure}[!t]
\includegraphics[width=\linewidth]{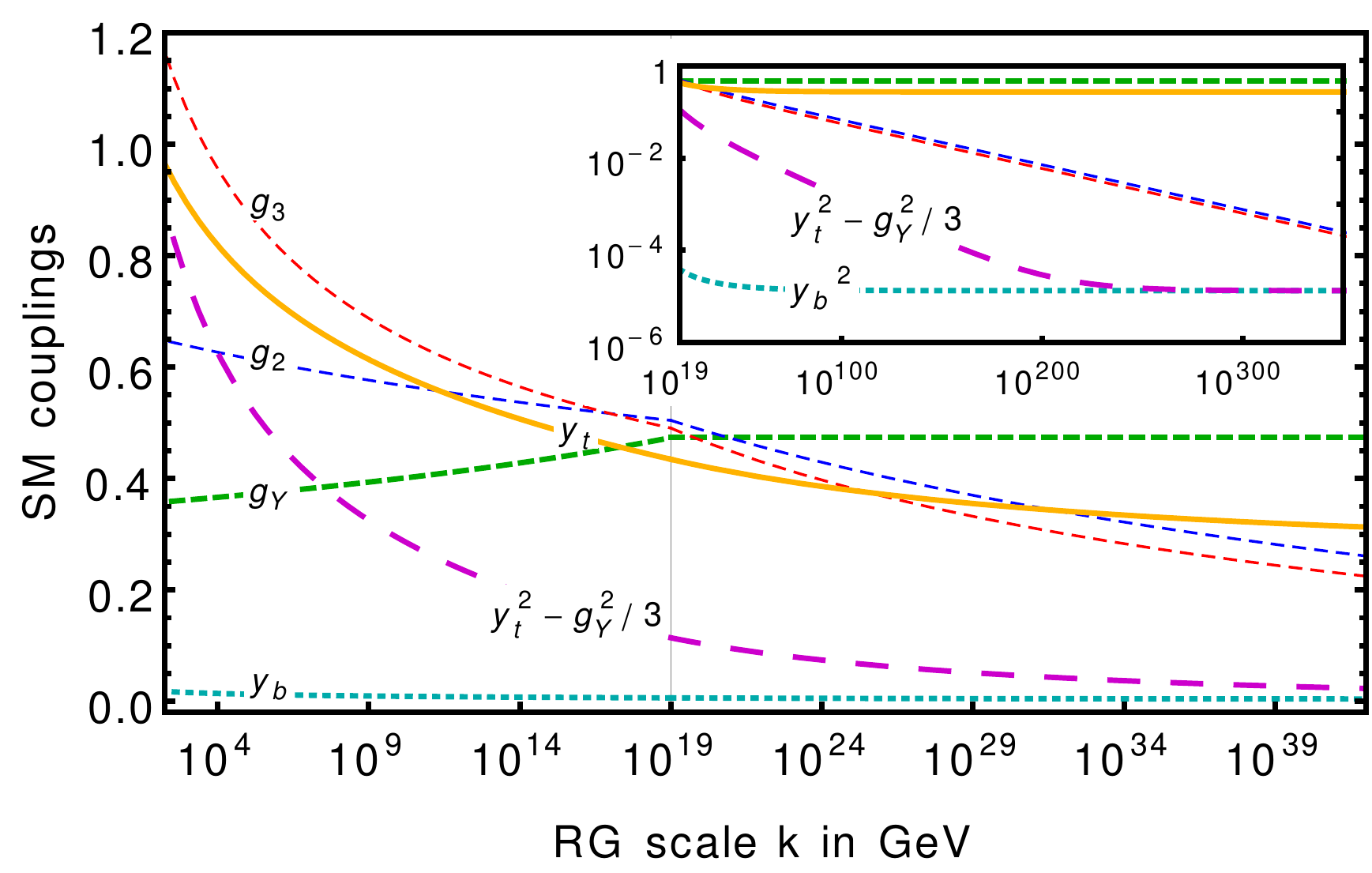}
\caption{\label{fig:flowplot} 
RG trajectory of Standard-Model couplings for $f_g=9.7\times10^{-3}$ and $f_y=1.188\times 10^{-4}$, reaching $g(k_{\rm IR})=0.358$, $y_t(k_{\rm IR})=0.965$, and $y_b(k_{\rm IR})= 0.018$ at $k_{\rm IR}=173\,\text{GeV}$. We also plot $y_t^2 - g_Y^2/3$ (pink, wide-dashed), which approaches $y_{b\,\ast}^2$ (dotted) in the far UV, cf.~Eq.~\eqref{eq:fprelation}.
}
\end{figure}
\begin{figure}[!t]
\includegraphics[width=\linewidth]{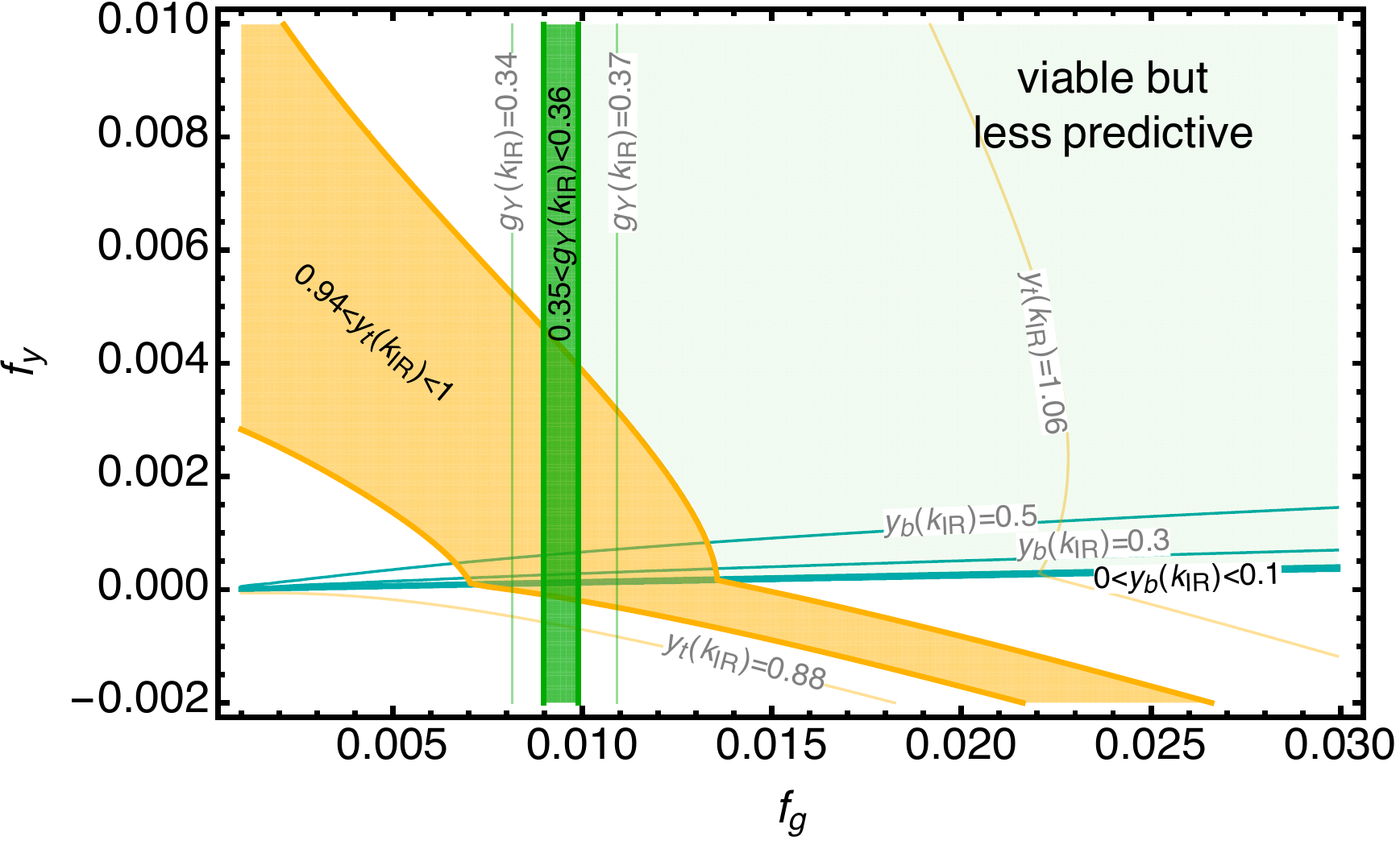}
\vspace{-17pt}
\caption{\label{fig:fyfgplot}
IR values of retrodicted couplings $g_Y(k_\text{IR})$, $y_t(k_\text{IR})$ and $y_b(k_\text{IR})$ at $k_\text{IR}=173\;\text{GeV}$ as a function of the two independent quantum-gravity contributions $f_g$ and $f_y$.
}
\end{figure}

The top quark is substantially heavier than all the other quarks, with a pole mass of $M_t\approx173\, \rm GeV$ \cite{ATLAS:2014wva} significantly larger than the pole mass of the second-heaviest quark, the bottom at $M_b\approx4.9\, \rm GeV$ \cite{Patrignani:2016xqp}. In the Standard Model, neither the two values nor their difference can be derived. The masses are determined by the Yukawa couplings $y_t, y_b$ to the Higgs, once it acquires a vacuum expectation value. The low-energy values of $y_t, y_b$ are free parameters in the Standard Model, fixed by comparing to experiment. We propose a mechanism that could generate the mass difference dynamically and uniquely determine the values of both masses from first principles. The mechanism follows from microscopic physics in the ultraviolet (UV), where an interplay of quantum gravity and gauge boson dynamics generates asymptotic safety \cite{Weinberg:1980gg, Reuter:1996cp}, i.e., an interacting Renormalization Group (RG) fixed point at transplanckian scales. This fixed point prevents Landau-pole type behavior in the running couplings, rendering the Standard Model UV-complete. The fixed point determines the values of $y_t$ and $y_b$ in the UV. This mechanism combines the fixed-point scenarios explored in \cite{Eichhorn:2017ylw,Eichhorn:2017lry}, see also \cite{Harst:2011zx}, where the top pole mass and Abelian gauge coupling are retrodicted separately. Due to the two quarks' unequal electric charges, $y_t$ and $y_b$ assume uniquely determined, different values at $M_{\rm Planck}=10^{19}\, \rm GeV$, cf.~Fig.~\ref{fig:flowplot}. This results in a retrodiction of unequal top and bottom masses at the electroweak scale. The viability of this mechanism hinges on the quantum numbers of the top and bottom quark: in our approximation, significant deviations from the measured charge ratio are incompatible with the observed masses.
\\
We now explain the mechanism, by following the RG flow from the UV fixed point through the transplanckian regime down to the electroweak scale.
\\
\indent\emph{Ultraviolet fixed point:}
There are strong indications for an asymptotically safe regime in quantum gravity, where the running gravitational couplings reach a scale-invariant regime that UV-completes the theory
\cite{Reuter:1996cp, Reuter:2001ag, Reuter:2004nx,
	Litim:2003vp, Codello:2008vh, Benedetti:2009rx, 
	Narain:2009fy, Manrique:2011jc, 
	Falls:2013bv, Dona:2013qba, Becker:2014qya, Meibohm:2015twa, 
	Gies:2016con, Eichhorn:2016vvy, Denz:2016qks, Gonzalez-Martin:2017gza, 
	Knorr:2017fus, Christiansen:2017bsy, Falls:2017lst}.
Quantum-gravity fluctuations impact the scale dependence of running matter couplings
\cite{Robinson:2005fj,Rodigast:2009zj,
	Daum:2009dn,Harst:2011zx,Folkerts:2011jz,Eichhorn:2017lry,
	Christiansen:2017gtg,Christiansen:2017cxa,
	Shaposhnikov:2009pv,
	Zanusso:2009bs,Oda:2015sma,Eichhorn:2017eht,Eichhorn:2016esv,
	Hamada:2017rvn,Eichhorn:2017ylw}. 
For the gauge couplings of the Standard Model, $g_3$ for SU(3), $g_2$ for SU(2) and $g_Y$ for the Abelian hypercharge, the one-loop beta functions and coefficients read \cite{GellMann:1954fq, Gross:1973id, Politzer:1973fx}
\bea
\beta_{g_i} &=& k \partial_k\, g_i(k)= b_{0,\,i}\,g_i^3/(16\pi^2)- f_g\, g_i, \label{eq:betasgauge}\\
b_{0,\, 3}&=&-7,\, b_{0,\,2}=-\frac{19}{6},\, b_{0, \, Y}= \frac{19+36 \left(Y_b^2+2Y_Q^2+Y_t^2\right)}{6}.\nonumber
\eea
$Y_{t,b,Q}$ are the hypercharges of the right-handed top and bottom quark and the left-handed SU(2) quark doublet, respectively.
$f_g$ encodes the quantum-gravity contribution that acts like an anomalous dimension for the gauge couplings, and we assume that additional terms are subleading. 
 These additional contributions are proportional to the product of $g_i$ and quantum-gravity-induced higher-order couplings. The
fixed-point values of the latter are of the same order as $f_g$, see the discussion in \cite{Eichhorn:2016esv,Christiansen:2017gtg,Eichhorn:2017eht}. They enter the $\beta_{g_i}$ through a loop diagram, leading to a suppression by $\frac{1}{16\pi^2}$ in comparison to the direct contribution in Eq.~\eqref{eq:betasgauge}, see \cite{Eichhorn:2017eht}.
We work with the one-loop beta functions to explain the mechanism, explicitly checking that two-loop effects only lead to quantitative changes. We focus on $f_g \geq 0$, as found in truncations of the functional RG flow \cite{Wetterich:1992yh, Morris:1993qb} under the impact of asymptotically safe quantum gravity \cite{Daum:2009dn,Harst:2011zx,Folkerts:2011jz,Eichhorn:2017lry,Christiansen:2017gtg,Christiansen:2017cxa}, see \cite{Niedermaier:2006wt,Reuter:2012id,Percacci:2017fkn,Eichhorn:2017egq} for reviews. In the asymptotically safe regime beyond the Planck scale, $f_g=\rm const.$ holds as a consequence of gravitational fixed-point scaling. For the non-Abelian gauge couplings, this reinforces the asymptotically free fixed point at $g_{3\,\ast}=0=g_{2\,\ast}$. For the Abelian gauge coupling, the positive one-loop coefficient, generated by screening quantum fluctuations of charged matter, and the antiscreening gravity contribution cancel at an interacting fixed point \cite{Harst:2011zx,Eichhorn:2017lry,Eichhorn:2017muy},
\be
\beta_{g_Y}\Big|_{g_Y=g_{Y\, \ast}}=0, \quad \quad g_{Y\,\ast}^2= \frac{16\pi^2}{b_{0,\, Y}} f_g\label{eq:gYast}.
\ee
Quantum-gravity contributions to the running of the Yukawas supplement the one-loop beta functions \cite{Cheng:1973nv}
\bea
\label{eq:betays}
\beta_{y_{t \,(b)}} &=&
\frac{y_{t \,(b)}}{16\,\pi^2}\;\left(\frac{3 y_{b \,(t)}^2}{2}+ \frac{9 y_{t \,(b)}^2}{2}  - \frac{9}{4}g_2^2 -8 g_3^2 \right) \nonumber\\
&{}&- f_y\, y_{t \,(b)}  -\frac{3 y_{t \,(b)}}{16\,\pi^2}\left(Y_Q^2+Y_{t \,(b)}^2\right)g_Y^2.
\eea
For the quantum-gravity contribution, $f_y=\rm const$ holds in the asymptotically safe transplanckian regime \cite{Zanusso:2009bs,Oda:2015sma,Eichhorn:2017eht,Eichhorn:2016esv,Hamada:2017rvn,Eichhorn:2017ylw} generating an interacting fixed point at $y_{t,b, \, \ast}\neq 0$ through the interplay with Abelian fluctuations: At the fixed point at $g_{2\,\ast}=0=g_{3\,\ast}$ and $g_{Y\,\ast}$ in Eq.~\eqref{eq:gYast}, we obtain
\bea
y_{t/b\,\ast}^2&=&\frac{8}{3}\pi^2\left( f_y+\frac{3f_g \left(2Y_Q^2+3Y_{t/b}^2-Y_{b/t}^2\right)}{2\,b_{0,\, Y}}\right). \label{eq:fpys}
\eea
Specifying to Standard-Model charges $Y_t=2/3$, $Y_b=-1/3$, and $Y_Q=1/6$, yields  a fixed-point equation  that is the key relation of our scenario
\be
 y_{t\,\ast}^2-y_{b\,\ast}^2 = \frac{1}{3}g_{Y\, \ast}^2.\label{eq:fprelation}
\ee
This relation enforces $y_{t\,\ast} \neq y_{b\,\ast}$ in the far UV because $g_{Y\, \ast}\neq 0$.
The difference in fixed-point values, $y_{t(b)\,\ast}$, has an intuitive physical interpretation: The interacting fixed point for the Yukawas is generated through a balance of quantum fluctuations of matter with  gauge and gravity fluctuations. The two fixed-point values $y_{t(b)\, \ast}$ must be unequal since Abelian gauge boson fluctuations couple more strongly to the top than to the bottom quark, as the top has a larger hypercharge, i.e., $Y_t^2>Y_b^2$. To compensate the combined impact of gravity and gauge boson fluctuations and generate a fixed point, the top Yukawa coupling must be larger, $y_{t\,\ast}>y_{b\,\ast}$.
\\
The beta functions in Eqs.~\eqref{eq:betasgauge} and \eqref{eq:betays} admit further fixed-point solutions, e.g., $g_{Y\, \ast}=0$, $y_{b\, \ast}=0$, $y_{t\, \ast}>0$ explored in \cite{Eichhorn:2017ylw}, cf.~light-green shaded region in Fig.~\ref{fig:fyfgplot}. Here, we focus on the most predictive fixed-point solution, cf.~Eqs.~\eqref{eq:gYast} and \eqref{eq:fpys}, leading to retrodictions of the top mass $M_t$, the bottom mass $M_b$ and the Abelian hypercharge coupling $g_Y$ at the electroweak scale. 
\\
\indent\emph{RG flow at transplanckian scales:}
Starting from Eq.~\eqref{eq:fprelation}, the couplings deviate from their fixed-point values during the RG flow towards the infrared (IR). For real fixed-point values, Eq.~\eqref{eq:fprelation} implies $y_{t\, \ast}>y_{b\, \ast}$, and the RG flow conserves this inequality: The ratio $y_t(k)/y_b(k)$ cannot become smaller than 1 if $y_{t\,\ast}/y_{b\,\ast}>1$ in the UV. The flow of the ratio is given by
\be
\beta_{\frac{y_t}{y_b}}= \frac{1}{16\pi^2} \frac{y_t}{y_b} \left(3(y_t^2-y_b^2)- g_Y^2 \right).\label{eq:ratioflow}
\ee 
For $y_t(k)/y_b(k)\rightarrow1$ from above, the beta function becomes negative due to the contribution of the Abelian gauge coupling. Hence, the ratio $y_t(k)/y_b(k)$ is driven away from 1 towards larger values. Once created by the fixed-point structure, the relation $y_t(k)-y_b(k)>0$ is thus preserved down to the IR, cf.~Fig.~\ref{fig:flowplot}.
\\
Specifically, the trajectories in Fig.~\ref{fig:flowplot} arise as follows.
Since $f_g=\text{const.}$  in the transplanckian regime, $g_Y(k>M_{\rm Planck})=g_{Y\,\ast}$ holds. This results from the competition of the two distinct contributions in Eq.~\eqref{eq:betasgauge}: The screening matter contribution, encoded in $b_{0,\, Y}g_Y^3>0$ drives any small deviation $g_Y(k)= g_{Y\, \ast}+\delta$ with $\delta>0$ back to $\delta=0$ under the RG flow to the IR. Conversely, the antiscreening gravity contribution, encoded in $-f_g\, g_Y<0$, drives any small deviation $g_Y(k)= g_{Y\, \ast}-\delta$ with $\delta>0$ back to $\delta=0$. In other words, the fixed point is IR attractive, cf.~thick dashed green line in Fig.~\ref{fig:flowplot}.
\\
This is in contrast to the behavior of the non-Abelian gauge couplings, where the gravity contribution triggers a power-law running in the transplanckian regime. Since both, the gravity-contribution and the matter contribution to the beta functions $\beta_{g_{2,3}}$ are antiscreening,  the free fixed-point is IR repulsive. Hence, deviations from it are allowed in the transplanckian regime and $g_{2,3}$ grow under the RG flow to the IR, until they reach the experimentally determined values at IR scales.
\\
This dynamics  for the gauge couplings leads to a more intricate  behavior of the Yukawas: although the fixed point in Eq.~\eqref{eq:fpys} is IR attractive, the Yukawas run  as soon as $g_{2,3}$ deviate from zero significantly, cf.~Fig.~\ref{fig:flowplot}. Their running is determined by a critical trajectory $y_{t(b)}(k)= y_{t(b)}(g_2(k), g_3(k))$  on which they  exhibit a slight growth towards the IR. The non-Abelian gauge contribution to the flow of the Yukawas is negative. This counteracts the screening effect of matter fluctuations. Thus, tiny deviations $y_t(k)= y_{t\, \ast}+\delta$ with $\delta>0$ are no longer driven back exactly to $y_{t\, \ast}$ for $g_{2,3}(k)>0$. The critical trajectory is IR attractive, i.e., starting from their fixed-point values, the Yukawa couplings are fixed uniquely at $M_\text{Planck}$.
\\
\indent\emph{RG flow between the Planck and the electroweak scale:}
At the Planck scale, quantum-gravity effects switch off dynamically as $f_g$, $f_y$ are proportional to the Newton coupling measured in units of $k$. In asymptotic safety, it is constant at transplanckian scales, but falls off as $k^{-2}$ below $M_\text{Planck}$, making quantum-gravity effects negligible there, cf.~\cite{Reuter:2001ag, Reuter:2004nx}. 
To model this behavior, we implement a sharp transition to $f_g=0=f_y$ for $k\leq M_\text{Planck}$.  
Below $M_{\rm Planck}$, we follow the one-loop running in the Standard Model, attracted by a partial IR fixed-point \cite{Pendleton:1980as, Hill:1980sq, Wetterich:1981ir}. At the electroweak scale, where the Higgs acquires a vacuum expectation value, the two Yukawas determine the top and bottom mass. The inequality $y_t(k)>y_b(k)$, generated by the properties of the transplanckian regime, is preserved under the Standard-Model flow, as Eq.~\eqref{eq:ratioflow} still holds. The difference in fixed-point values between $y_t$ and $y_b$ thus generates a mass difference between $M_t$ and $M_b$.
\\
So far, we have explained how a mass difference between the two quarks could result from their unequal quantum numbers as a consequence of an asymptotically safe fixed point. We now test the quantitative viability of this mechanism in our approximation by using approximately observationally viable values.
To accommodate $g_Y(k_{\rm IR}=173\, {\rm GeV})=0.358$ in accordance with observations, $f_g= 9.7\times10^{-3}$ is required. Together with the values $g_2(k_{\rm IR})=0.64779$ and $g_3(k_{\rm IR})=1.1666$ see, e.g., \cite{Buttazzo:2013uya} this also fixes the running of the non-Abelian gauge couplings at all scales. Then, $f_y=1.188\cdot 10^{-4}$ is required to obtain $y_b(k=4.2\, \rm GeV)=0.024$. This translates into a bottom pole mass \cite{Patrignani:2016xqp} of $M_{b}=4.9\,\text{GeV}$.
Given this input, the mechanism presented here generates $y_t(k=168\,\rm GeV)=0.967$ corresponding to a top pole mass \cite{Patrignani:2016xqp} of $M_{t}=178\,\text{GeV}$. All three retrodicted quantities, $M_{t}$, $M_{b}$ and $g_Y$, come out rather close to their observed values  with the input of two free parameters, $f_y$ and $f_g$. The above values $f_y, f_g$ lie in the vicinity of fixed-point values obtained in an approximation for quantum gravity minimally coupled to matter fields of the Standard Model \cite{Dona:2013qba}. A quantitatively precise calculation of $f_y, f_g$ is subject to future studies. These studies must include higher-order curvature operators as in \cite{Hamada:2017rvn,Eichhorn:2017eht} and non-minimal matter-curvature couplings as in \cite{Narain:2009fy,Oda:2015sma,Eichhorn:2017sok} to determine the gravitational fixed-point values which directly set $f_g$ and $f_y$.
\\
As the UV fixed point is generated from a balance of the leading quantum-gravity contribution with the one-loop matter contribution and lies at small Standard-Model couplings, its existence is expected to be stable under the extension to higher-loop orders in the Standard-Model sector. 
Including two-loop terms in the Standard-Model running \cite{Caswell:1974gg, Jones:1974mm, Jones:1981we, Fischler:1982du, Machacek:1983tz, Machacek:1983fi}, $f_g=9.8\times10^{-3}$ 
yields $g_Y(k_{\rm IR})=0.358$ and $f_y=1.1266\times10^{-4}$ gives a bottom pole-mass of $M_{b}=4.9\,\text{GeV}$. This retrodicts a top pole-mass of $M_t =182\, \rm GeV$.
\\
Analyzing an extended setting going beyond the third generation could provide a future test of the present model.
Extending our study to the quarks of the second generation requires to account for the CKM mixing matrix. Inspecting
the  beta-functions for the strange and charm Yukawa under the simplifying assumption of a diagonal mixing matrix at $y_{t\,\ast}$, $y_{b\,\ast}$ and $g_{Y\,\ast}$ yields a fixed point at vanishing Yukawas for charm and strange which is IR-attractive in the strange and thus retrodicts $M_s/M_t\simeq 0$. Testing whether the tiny ratio $M_s/M_t\approx 5\cdot 10^{-4}$ is compatible with our setting requires to go beyond the above simplifying assumptions in more complete studies  but should provide a critical future test of the present proposal. In the charm, this fixed point is IR repulsive,  rendering the charm asymptotically free. Therefore $M_c/M_t$ is not retrodicted. Specifically, $M_c/M_t \approx7 \cdot 10^{-3}$ can be accommodated in our setting.
\\
\indent\emph{Exploring the gravitational parameter space:}
We now explore $f_g$ and $f_y$ away from the specific values used above. 
This exploits the link between electroweak and Planck-scale physics in order to constrain the microscopic gravitational parameter space by the requirement to match IR observables, in the spirit of \cite{Eichhorn:2017eht}.
\\
In our approximation, the low-energy value of $g_Y$ only depends on $f_g$. Hence, lines of constant $f_g$ in Fig.~\ref{fig:fyfgplot} correspond to lines of fixed $g_Y(k_{\rm IR}=173\, {\rm GeV})$. In contrast, $y_{t/b}(k_\text{IR})$ depend on $f_y$ as well as on $f_g$ through the gauge contributions in Eq.~\eqref{eq:betays}. Thus, lines of constant $y_{t/b}(k_\text{IR})$ are not simply lines of constant $f_y$.
\\
Fig.~\ref{fig:fyfgplot} visualizes that the existence of an intersection area of the three approximately observationally viable contours defined by $0<y_b(k_\text{IR})<0.1$, $0.94<y_t(k_\text{IR})<1$ and $0.35<g_Y(k_\text{IR})<0.36$ is a nontrivial result. 
An intersection does not occur for arbitrary combinations of values. For instance, $g_y(k_\text{IR})>0.4$ and $0.94<y_t(k_\text{IR})<1$ are incompatible with a non-zero bottom mass in our approximation. Thus, in our approximation, values close to the observed ones appear to be singled out by asymptotic safety.
\\
\begin{figure}[!t]
\vspace{-4pt}
\includegraphics[width=\linewidth]{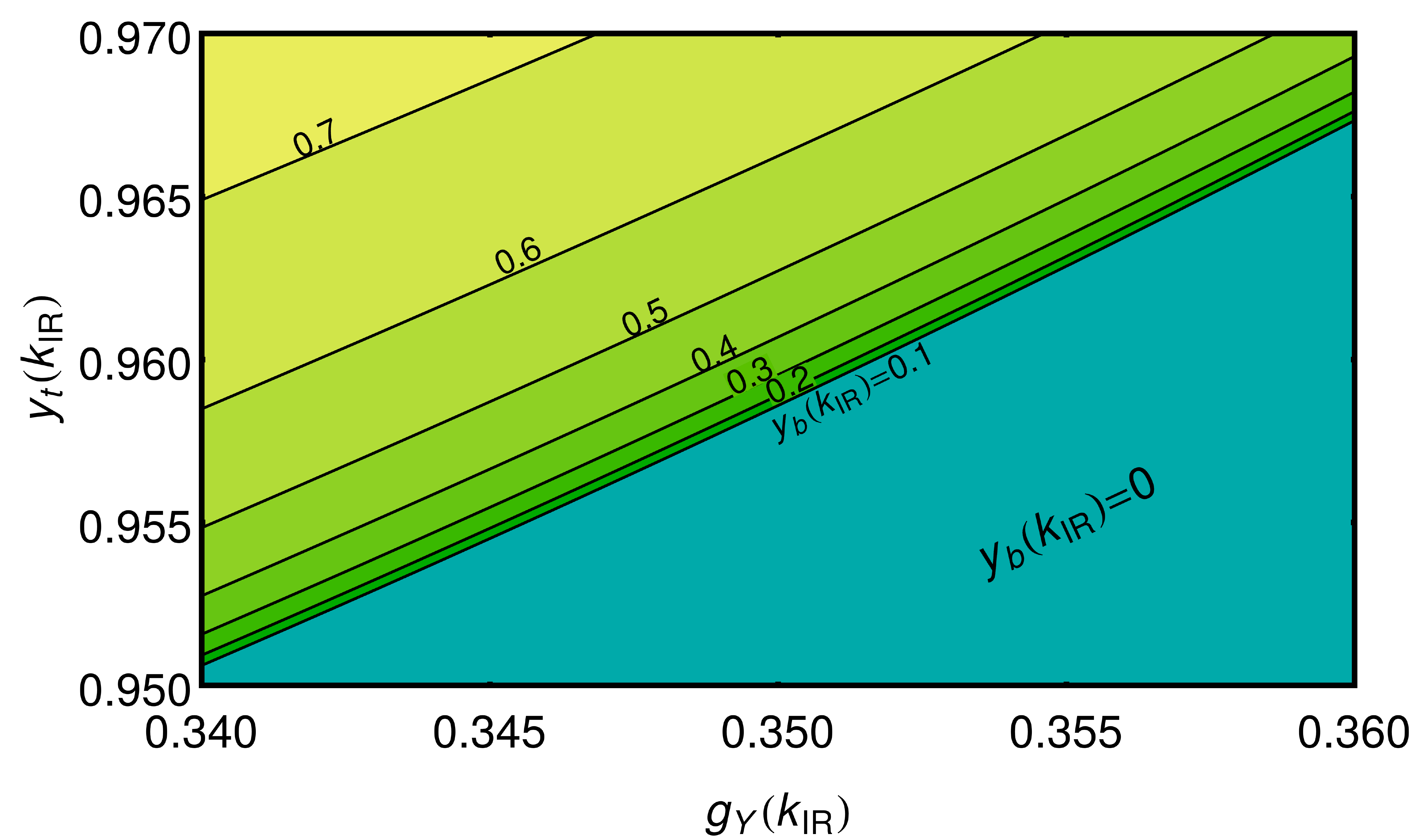}
\caption{\label{fig:criticality} 
Bottom Yukawa coupling $y_b(k_{\rm IR})$ at $k_{\rm IR}= 173\, \rm GeV$ as a function of the IR values of $g_Y(k_{\rm IR})$ and $y_t(k_{\rm IR})$.}
\end{figure}
The fixed-point in Eq.~\eqref{eq:fprelation} shows that $y_{b\, \ast}^2$ depends on the difference of the squares of $y_{t\, \ast}$ and $g_{Y\, \ast}$. Accordingly, small variations of these two numbers away from $y_{t\, \ast}^2= g_{Y\, \ast}^2/3$ result in a fast growth of the value of $y_b (k_{\rm IR})$. Due to the different U(1) hypercharges of top and bottom, the line $M_b=M_t$ cannot be reached, and a difference $M_t-M_b>0$ always persists. On the other hand, a very large difference, $M_t -M_b \simeq M_t$ requires a choice of the gravity parameters in a relatively small region of the gravitational parameter space, such that the system sits close to the phase-transition line to vanishing bottom mass.
In our approximation, this region translates into close-to Standard-Model values for $g_Y(k_{\rm IR})$ and $M_t$, cf.~Fig.~\ref{fig:criticality}.
\\
In summary, we have uncovered a non-trivial UV fixed point for the Standard Model couplings $g_{Y\,\ast}\neq 0$ and $y_{t(b)\,\ast}\neq 0$ induced by asymptotically safe gravity, that generically results in a mass difference between the top and bottom quark, i.e., $M_t>M_b$.
This fixed point retrodicts $(g_Y(k_{\rm IR}),\,M_t,\,M_b)$, in terms of two gravitational parameters $(f_g,\,f_y)$. In our study the retrodiction is in approximate agreement with the observed IR values, cf.~Fig.~\ref{fig:fyfgplot}.
\begin{figure}[!t]
\vspace*{-6pt}
\includegraphics[width=\linewidth]{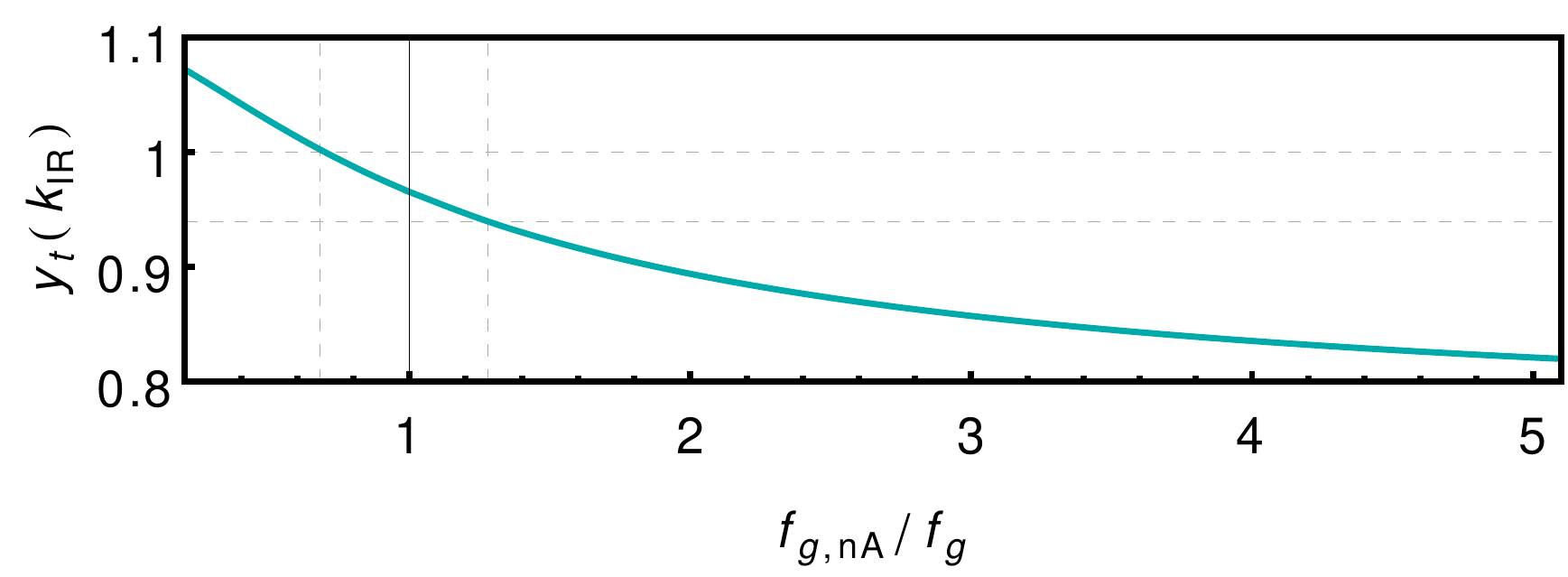}\\
\includegraphics[width=\linewidth]{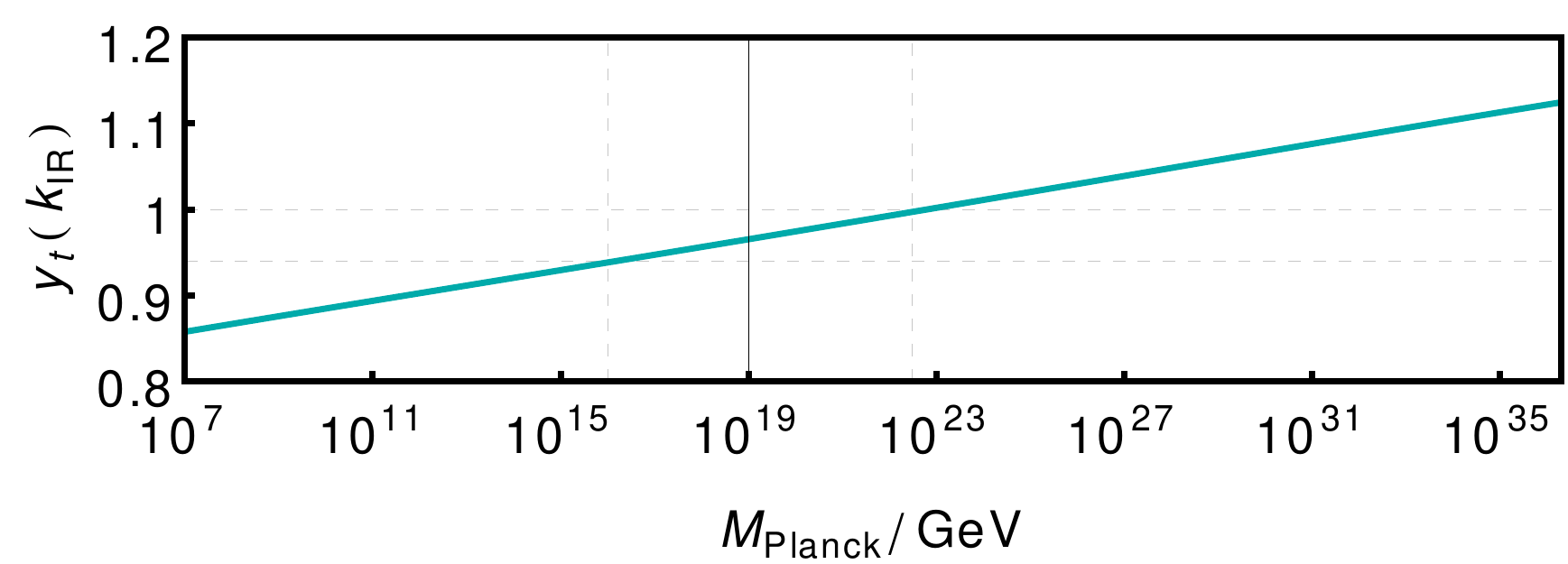}\\
\caption{\label{fig:testdeviations}
Top-Yukawa coupling $y_t(k_{\rm IR})$ at $k_\text{IR}=173\,\rm GeV$ as a function of a non-universal gravity contribution $f_{g,\text{nA}}/f_g$ (upper panel) and of a modified Planck scale $M_\text{Planck}$ (lower panel) for fixed $g_Y (k_{\rm IR})=0.358$ and $M_b=4.9\, \rm GeV$.
}
\end{figure}
\\
\indent\emph{Three observations: 
\\
1) Universality of gravity contributions:}
A key assumption of our study is the independence of the quantum-gravity contributions from internal symmetries:
gravity is the only known force that couples universally to all matter fields such that $f_g$ is independent of the gauge group.  A significant violation of this universality leads to a quantitative failure of the above scenario. Specifically, let the gravitational contribution to the running of the non-Abelian gauge couplings be given by $f_g \rightarrow f_{g, \, nA}$ in Eq.~\eqref{eq:betasgauge}. The rate at which $g_{2,3}$ grow above the Planck scale is thereby increased (lowered) for $f_{g,\, nA}>(<)f_g$. This affects how fast the Yukawa couplings increase in the transplanckian regime.  Only $f_{g, \, nA} \approx f_g$ results in an observationally viable range for $y_t(k_{\rm IR})$, cf.~Fig.~\ref{fig:testdeviations}. Thus, the independence of the gravitational contribution from the gauge group is suggested by the observed values of $y_{b}(k_{\rm IR})$, $y_{t}(k_{\rm IR})$ and $g_Y(k_{\rm IR})$.
\\
\emph {2) Setting the scale:}
A second central assumption underlying our study is that the scale at which the gravitational contributions switch off is the Planck scale. We test whether another, presently unknown universally coupled interaction could underlie the proposed mechanism. Its scale would of course not be tied to the Planck scale.
Varying the scale significantly away from $10^{19}\, \rm GeV$ results in a mismatch of $M_b/M_t$ with the observed values, cf.~upper panel in  Fig.~\ref{fig:testdeviations}. Given the electroweak scale, which is an input of our calculation, the Planck mass can thus be estimated by demanding that the model realizes a mass ratio in the vicinity of the observed ratio of $M_b/M_t$ in our approximation.
\begin{figure}[!t]
\includegraphics[width=\linewidth]{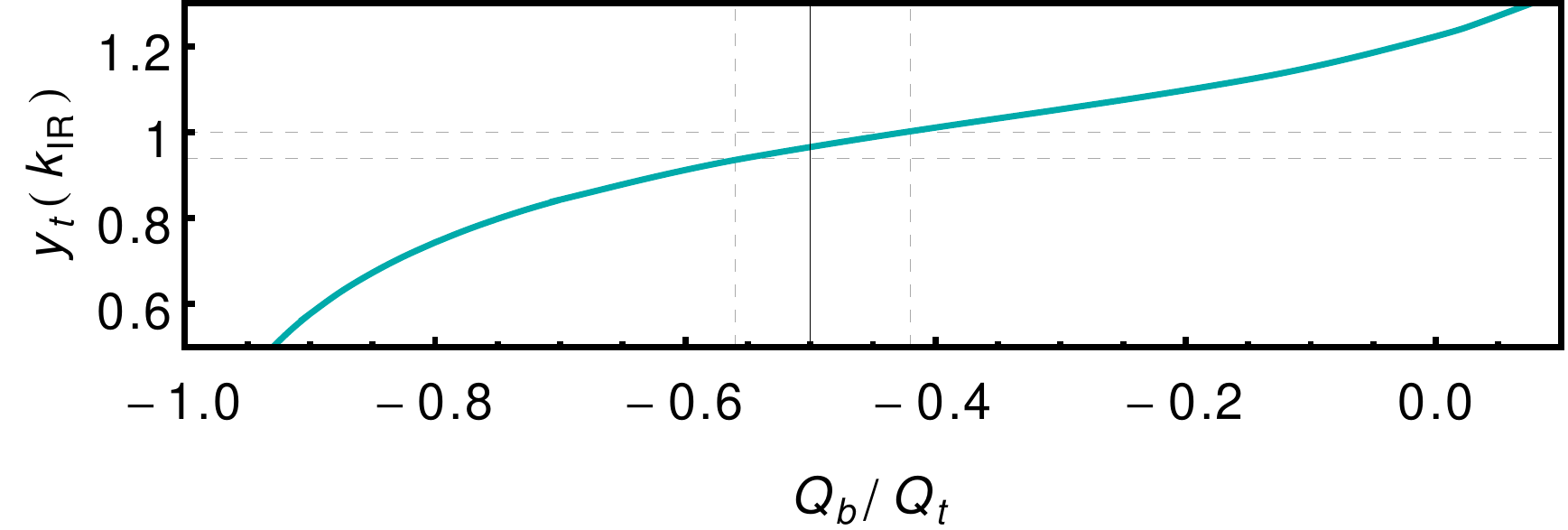}
\caption{\label{fig:chargeratio} Top-Yukawa coupling $y_t(k_{\rm IR})$ at $k_\text{IR}=173\,\rm GeV$ as a function of the charge ratio $Q_b/Q_t$ for fixed $g_Y (k_{\rm IR})=0.358$ and $M_b=4.9\, \rm GeV$.}
\end{figure}
\begin{figure}[!t]
\includegraphics[width=0.45\linewidth]{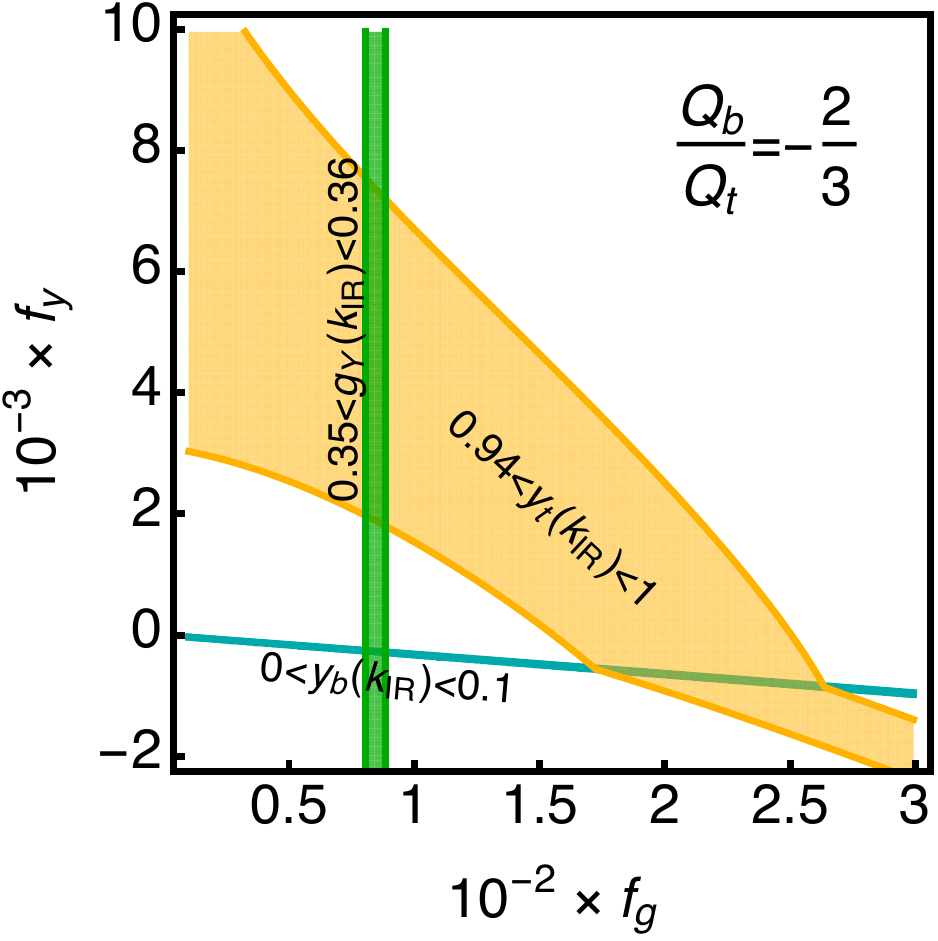}\quad\quad \includegraphics[width=0.45\linewidth]{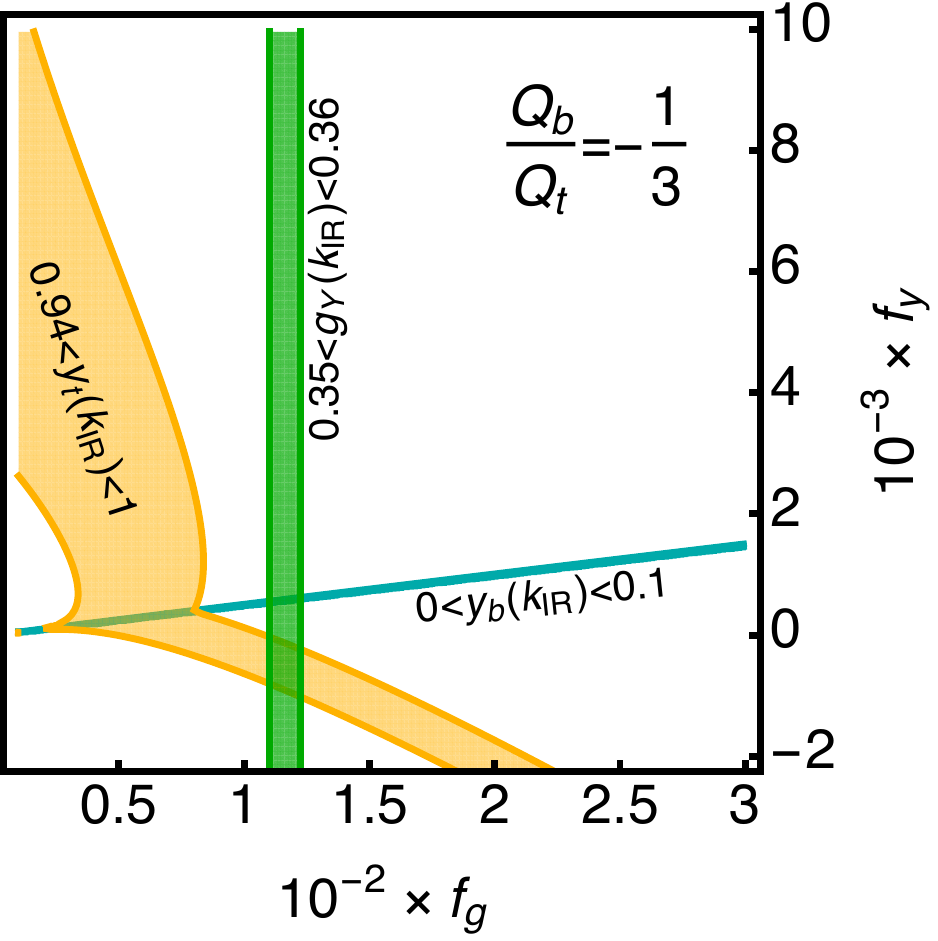}
\caption{\label{fig:chargeratiofyfg}
IR-values of the retrodicted couplings $g_Y(k_{\rm IR})$, $y_t(k_{\rm IR})$ and $y_b(k_{\rm IR})$ at $k_\text{IR}=173\,\text{GeV}$ as a function of the two quantum gravity contributions $f_g$ and $f_y$ at modified charge ratio, $Q_b/Q_t=-2/3$ (left-hand panel); $Q_b/Q_t=-1/3$ (right-hand panel).
}
\end{figure}
\\
\emph{ 3) Selecting electric charges for top and bottom:}
The top-bottom mass-difference is rooted in distinct fixed-point values in Eq.~\eqref{eq:fprelation}. Varying the quantum numbers of the top and bottom from their values in the Standard Model results in a modified running of $g_Y,\, y_t$ and $y_b$ and an altered fixed-point relation
\be
\label{eq:fpConditionCharges}
y_{b\, \ast}^2= y_{t\,\ast}^2-(Q_t^2 - Q_b^2)g_{Y\,\ast}^2\;.
\ee
Here, we keep the top and bottom in a doublet of the SU(2). The hypercharges of the doublet $Y_Q$ and singlets $Y_{b/t}$ are linked to the electric charges by $Y_t = Q_t,\, Y_b = Q_b = Q_t-1,\, Y_Q = Q_t-\frac{1}{2}$, where the last equality ensures equal electric charges for the right- and left-handed quarks.
It turns out that for $Q_b/Q_t<-1/2$, $M_t/M_b \rightarrow 0$, whereas for $Q_b/Q_t>-1/2$, $M_t/M_b \rightarrow 1$, cf.~Fig.~\ref{fig:chargeratio}. The reason lies in the dynamics of the green, cyan and yellow contours in Fig.~\ref{fig:chargeratiofyfg}: An increase in $Q_b/Q_t$ triggers a growth in $f_g$, since $b_{0,Y}$ increases with $Q_b/Q_t$. Thus, the green contour moves to the right as a function of $Q_b/Q_t$. Simultaneously, the cyan and yellow contours move towards each other as $y_{b\, \ast}\rightarrow y_{t\, \ast}$ for $Q_b/Q_t\rightarrow 1$. Accordingly, the three contours single out a value of $Q_b/Q_t$ at which they intersect in one location in the $f_g, f_y$ plane. This value agrees with the Standard Model value $Q_b/Q_t=-1/2$.
\\
\noindent\emph{Conclusions:}
The asymptotic-safety paradigm could provide a UV completion for quantum gravity coupled to the Standard Model. At an asymptotically safe fixed point, residual interactions in the microscopic regime can imprint a nontrivial structure on the low-energy masses of the model. Thereby, observations such as $M_t\gg M_b$ could become an automatic consequence of the asymptotically safe regime. Our study hints at the potential predictive power of an asymptotically safe UV regime. The mechanism we propose here links the measured ratio of electric charges of top and bottom to their masses: If the charge ratio deviates significantly from the Standard Model value, in our setting no choice of microscopic gravitational parameters is available to correctly retrodict $M_t$, $M_b$ and $g_Y$.
\\
\emph{Acknowledgements:} We acknowledge helpful comments from A.~Pereira. This research is supported by an Emmy-Noether grant of the DFG under grant no. EI/1037-1. A.~Held acknowledges support by the Studienstiftung des deutschen Volkes. This research was supported in part by Perimeter Institute for Theoretical Physics. Research at Perimeter Institute is supported by the Government of Canada through the Department of Innovation, Science, and Economic Development, and by the Province of Ontario through the Ministry of Research and Innovation.


\begin{thebibliography}{99}
%

\bibitem{ATLAS:2014wva} 
  [ATLAS and CDF and CMS and D0 Collaborations],
  arXiv:1403.4427 [hep-ex].
  
\bibitem{Patrignani:2016xqp} 
  C.~Patrignani {\it et al.} [Particle Data Group],
  Chin.\ Phys.\ C {\bf 40}, no. 10, 100001 (2016).


\bibitem{Weinberg:1980gg} 
  S.~Weinberg,
{\it  In *Hawking, S.W., Israel, W.: General Relativity*, 790-831}
(Cambridge University Press, Cambridge, 1980).

\bibitem{Reuter:1996cp} 
  M.~Reuter,
  Phys.\ Rev.\ D {\bf 57}, 971 (1998).
  
  
    
\bibitem{Eichhorn:2017ylw} 
  A.~Eichhorn and A.~Held,
  Phys.\ Lett.\ B {\bf 777}, 217 (2018).
  
\bibitem{Eichhorn:2017lry} 
  A.~Eichhorn and F.~Versteegen,
  JHEP {\bf 1801}, 030 (2018).
  
\bibitem{Harst:2011zx} 
  U.~Harst and M.~Reuter,
  JHEP {\bf 1105}, 119 (2011).

\bibitem{Reuter:2001ag} 
  M.~Reuter and F.~Saueressig,
  Phys.\ Rev.\ D {\bf 65}, 065016 (2002).

\bibitem{Litim:2003vp} 
  D.~F.~Litim,
  Phys.\ Rev.\ Lett.\  {\bf 92}, 201301 (2004).
  
\bibitem{Reuter:2004nx} 
  M.~Reuter and H.~Weyer,
  JCAP {\bf 0412}, 001 (2004).

\bibitem{Codello:2008vh} 
  A.~Codello, R.~Percacci and C.~Rahmede,
  Annals Phys.\  {\bf 324}, 414 (2009).
  
\bibitem{Benedetti:2009rx} 
  D.~Benedetti, P.~F.~Machado and F.~Saueressig,
  Mod.\ Phys.\ Lett.\ A {\bf 24}, 2233 (2009).

\bibitem{Narain:2009fy} 
  G.~Narain and R.~Percacci,
  Class.\ Quant.\ Grav.\  {\bf 27}, 075001 (2010).
  
\bibitem{Manrique:2011jc} 
  E.~Manrique, S.~Rechenberger and F.~Saueressig,
  Phys.\ Rev.\ Lett.\  {\bf 106}, 251302 (2011).
 
\bibitem{Falls:2013bv} 
  K.~Falls, D.~F.~Litim, K.~Nikolakopoulos and C.~Rahmede,
  arXiv:1301.4191 [hep-th].

\bibitem{Dona:2013qba} 
  P.~Don\`a , A.~Eichhorn and R.~Percacci,
  Phys.\ Rev.\ D {\bf 89}, no. 8, 084035 (2014).
  
\bibitem{Becker:2014qya} 
  D.~Becker and M.~Reuter,
  Annals Phys.\  {\bf 350}, 225 (2014).
  
\bibitem{Meibohm:2015twa} 
  J.~Meibohm, J.~M.~Pawlowski and M.~Reichert,
  Phys.\ Rev.\ D {\bf 93}, no. 8, 084035 (2016).
  
\bibitem{Gies:2016con} 
  H.~Gies, B.~Knorr, S.~Lippoldt and F.~Saueressig,
  Phys.\ Rev.\ Lett.\  {\bf 116}, no. 21, 211302 (2016).
  
\bibitem{Eichhorn:2016vvy} 
  A.~Eichhorn and S.~Lippoldt,
  Phys.\ Lett.\ B {\bf 767}, 142 (2017).
  
\bibitem{Denz:2016qks} 
  T.~Denz, J.~M.~Pawlowski and M.~Reichert,
  arXiv:1612.07315 [hep-th].
  
\bibitem{Knorr:2017fus} 
  B.~Knorr and S.~Lippoldt,
  Phys.\ Rev.\ D {\bf 96}, no. 6, 065020 (2017).
    
\bibitem{Gonzalez-Martin:2017gza} 
  S.~Gonzalez-Martin, T.~R.~Morris and Z.~H.~Slade,
  Phys.\ Rev.\ D {\bf 95}, no. 10, 106010 (2017).
  
\bibitem{Christiansen:2017bsy} 
  N.~Christiansen, K.~Falls, J.~M.~Pawlowski and M.~Reichert,
  Phys.\ Rev.\ D {\bf 97}, no. 4, 046007 (2018).
  
\bibitem{Falls:2017lst} 
  K.~Falls, C.~S.~King, D.~F.~Litim, K.~Nikolakopoulos and C.~Rahmede,
  arXiv:1801.00162 [hep-th].
  
\bibitem{Robinson:2005fj} 
  S.~P.~Robinson and F.~Wilczek,
  Phys.\ Rev.\ Lett.\  {\bf 96}, 231601 (2006).

\bibitem{Rodigast:2009zj} 
  A.~Rodigast and T.~Schuster,
  Phys.\ Rev.\ Lett.\  {\bf 104}, 081301 (2010).
  
\bibitem{Daum:2009dn} 
  J.~E.~Daum, U.~Harst and M.~Reuter,
  JHEP {\bf 1001}, 084 (2010).
  
\bibitem{Folkerts:2011jz} 
  S.~Folkerts, D.~F.~Litim and J.~M.~Pawlowski,
  Phys.\ Lett.\ B {\bf 709}, 234 (2012).
  
\bibitem{Christiansen:2017gtg} 
  N.~Christiansen and A.~Eichhorn,
  Phys.\ Lett.\ B {\bf 770}, 154 (2017).
  
\bibitem{Christiansen:2017cxa} 
  N.~Christiansen, D.~F.~Litim, J.~M.~Pawlowski and M.~Reichert,
  arXiv:1710.04669 [hep-th].

\bibitem{Shaposhnikov:2009pv} 
  M.~Shaposhnikov and C.~Wetterich,
  Phys.\ Lett.\ B {\bf 683}, 196 (2010).
  
\bibitem{Zanusso:2009bs} 
  O.~Zanusso, L.~Zambelli, G.~P.~Vacca and R.~Percacci,
  Phys.\ Lett.\ B {\bf 689}, 90 (2010).
  
\bibitem{Oda:2015sma} 
  K.~y.~Oda and M.~Yamada,
  Class.\ Quant.\ Grav.\  {\bf 33}, no. 12, 125011 (2016).
  
\bibitem{Eichhorn:2016esv} 
  A.~Eichhorn, A.~Held and J.~M.~Pawlowski,
  Phys.\ Rev.\ D {\bf 94}, no. 10, 104027 (2016).

\bibitem{Eichhorn:2017eht} 
  A.~Eichhorn and A.~Held,
  Phys.\ Rev.\ D {\bf 96}, no. 8, 086025 (2017).
  
\bibitem{Hamada:2017rvn} 
  Y.~Hamada and M.~Yamada,
  JHEP {\bf 1708}, 070 (2017).
    
\bibitem{GellMann:1954fq} 
  M.~Gell-Mann and F.~E.~Low,
  Phys.\ Rev.\  {\bf 95}, 1300 (1954).

\bibitem{Gross:1973id} 
  D.~J.~Gross and F.~Wilczek,
  Phys.\ Rev.\ Lett.\  {\bf 30}, 1343 (1973).
  
\bibitem{Politzer:1973fx} 
  H.~D.~Politzer,
  Phys.\ Rev.\ Lett.\  {\bf 30}, 1346 (1973).

\bibitem{Wetterich:1992yh} 
  C.~Wetterich,
  Phys.\ Lett.\ B {\bf 301}, 90 (1993).
  
\bibitem{Morris:1993qb} 
  T.~R.~Morris,
  Int.\ J.\ Mod.\ Phys.\ A {\bf 9}, 2411 (1994).

\bibitem{Niedermaier:2006wt} 
  M.~Niedermaier and M.~Reuter,
  Living Rev.\ Rel.\  {\bf 9}, 5 (2006).
  
\bibitem{Reuter:2012id} 
  M.~Reuter and F.~Saueressig,
  New J.\ Phys.\  {\bf 14}, 055022 (2012).
  
\bibitem{Percacci:2017fkn} 
  R.~Percacci,
  ``An Introduction to Covariant Quantum Gravity and Asymptotic Safety,''
  World Scientific Publishing (2017).
  
\bibitem{Eichhorn:2017egq} 
  A.~Eichhorn,
  arXiv:1709.03696 [gr-qc].
  
\bibitem{Eichhorn:2017muy} 
  A.~Eichhorn, A.~Held and C.~Wetterich,
  arXiv:1711.02949 [hep-th].
 
\bibitem{Cheng:1973nv} 
  T.~P.~Cheng, E.~Eichten and L.~F.~Li,
  Phys.\ Rev.\ D {\bf 9}, 2259 (1974).

\bibitem{Buttazzo:2013uya} 
  D.~Buttazzo, G.~Degrassi, P.~P.~Giardino, G.~F.~Giudice, F.~Sala, A.~Salvio and A.~Strumia,
  JHEP {\bf 1312}, 089 (2013).
  
\bibitem{Eichhorn:2017sok} 
  A.~Eichhorn, S.~Lippoldt and V.~Skrinjar,
  Phys.\ Rev.\ D {\bf 97}, no. 2, 026002 (2018)
  doi:10.1103/PhysRevD.97.026002
  [arXiv:1710.03005 [hep-th]].

\bibitem{Caswell:1974gg} 
  W.~E.~Caswell,
  Phys.\ Rev.\ Lett.\  {\bf 33}, 244 (1974).

\bibitem{Jones:1974mm} 
  D.~R.~T.~Jones,
  Nucl.\ Phys.\ B {\bf 75}, 531 (1974).
  
\bibitem{Jones:1981we} 
  D.~R.~T.~Jones,
  Phys.\ Rev.\ D {\bf 25}, 581 (1982).
  
\bibitem{Fischler:1982du} 
  M.~Fischler and J.~Oliensis,
  Phys.\ Lett.\  {\bf 119B}, 385 (1982).
  
\bibitem{Machacek:1983tz} 
  M.~E.~Machacek and M.~T.~Vaughn,
  Nucl.\ Phys.\ B {\bf 222}, 83 (1983).
  
\bibitem{Machacek:1983fi} 
  M.~E.~Machacek and M.~T.~Vaughn,
  Nucl.\ Phys.\ B {\bf 236}, 221 (1984).

\bibitem{Pendleton:1980as} 
  B.~Pendleton and G.~G.~Ross,
  Phys.\ Lett.\  {\bf 98B}, 291 (1981).
  
\bibitem{Hill:1980sq} 
  C.~T.~Hill,
  Phys.\ Rev.\ D {\bf 24}, 691 (1981).
  
\bibitem{Wetterich:1981ir} 
  C.~Wetterich,
  Phys.\ Lett.\  {\bf 104B}, 269 (1981).
  
\end{thebibliography}
\end{document}